\documentclass[useAMS,usenatbib, letters]{mn2e}
\RequirePackage{lineno}
\usepackage{graphicx}  
\usepackage{ amssymb }
\usepackage{journals}
\usepackage[usenames]{color}
\def\simgt{\lower.5ex\hbox{$\; \buildrel > \over \sim \;$}}
\def\simlt{\lower.5ex\hbox{$\; \buildrel < \over \sim \;$}}

\def\Teff{$T_{\rm eff}$}

\def\alfa2d{MLT--$\,\alpha^{\rm 2D}$}

\definecolor{verdone}{rgb}{0,0.398,0}
\definecolor{stateblue}{rgb}{0.14, 0.0937, 0.508}

\input epsf

\usepackage{graphicx}
\usepackage{txfonts}
\usepackage{epsfig}
\usepackage[T1]{fontenc}
\usepackage{aecompl}

\newcommand{\msun}{\mbox{${\rm M}_{\odot}$}}
\title[Detection of solar-like oscillations in M4]{Detection of solar-like oscillations in relics of the Milky Way: asteroseismology of K giants in M4 using data from the NASA K2 mission}

\author[A. Miglio et al.]{A. Miglio$^{1,2}$\thanks{E-mail: miglioa@bison.ph.bham.ac.uk}, 
W. J. Chaplin$^{1,2}$, K. Brogaard$^{2}$, M. N. Lund$^{1,2}$, B. Mosser$^{3}$, G. R. Davies$^{1,2}$,  \newauthor R. Handberg$^{2}$, 
 A. P. Milone$^{4}$,  A. F. Marino$^{4}$, D. Bossini$^{1,2}$,  Y. P. Elsworth$^{1,2}$, \newauthor F. Grundahl$^{2}$,  T. Arentoft$^{2}$, L. R. Bedin$^{5}$, T. L. Campante$^{1,2}$,  J. Jessen-Hansen$^{2}$, \newauthor  C. D. Jones$^{1,2}$, J.  S. Kuszlewicz$^{1,2}$, L. Malavolta$^{5,6}$, V. Nascimbeni$^{5}$, E. L. Sandquist$^{7}$ \\
\\
$^{1}$School of Physics and Astronomy, University of Birmingham, Edgbaston, Birmingham B15 2TT, United Kingdom\\
$^{2}$Stellar Astrophysics Centre, Department of Physics and Astronomy, Aarhus University,  DK-8000 Aarhus C, Denmark \\
$^{3}$LESIA, Observatoire de Paris, PSL Research University, CNRS, Universit\'e Pierre et Marie Curie, Universit\'e Paris Diderot, 92195 Meudon, France\\
$^{4}$Research School of Astronomy \& Astrophysics, Australian National University, Mt Stromlo Observatory, ACT 2611, Australia \\
$^{5}$INAF - Osservatorio Astronomico di Padova, vicolo dell'Osservatorio 5, I-35122 Padova, Italy\\
$^{6}$Dipartimento di Fisica e Astronomia 'Galileo Galilei', Universit\'a di Padova, vicolo dell'Osservatorio 3, I-35122 Padova, Italy\\
$^{7}$Department of Astronomy, San Diego State University, USA\\
}     
\begin{document}
\date{}
\pagerange{\pageref{firstpage}--\pageref{lastpage}} \pubyear{2010}

\maketitle

\label{firstpage}
\vspace{-2cm}
\begin{abstract}
Asteroseismic constraints on K giants make it possible to infer radii,
masses and ages of tens of thousands of field stars. Tests against
independent estimates of these properties are however scarce,
especially in the metal-poor regime.  Here, we report the detection of
solar-like oscillations in 8 stars belonging to the red-giant branch
and red-horizontal branch of the globular cluster M4. The detections
were made in photometric observations from the K2 Mission during
its Campaign\,2.  
{Making use of independent constraints on the
distance, we estimate masses of the 8 stars by utilising different combinations of
seismic and non-seismic inputs. When introducing a correction to the $\Delta\nu$ scaling relation as suggested by stellar models,  for RGB stars we find excellent agreement with the expected masses from isochrone fitting, and
with a distance modulus derived using independent methods. The offset with respect to independent masses is lower, or comparable with, the uncertainties on the average RGB mass ($4-10\%$, depending on the combination of constraints used).}
Our results
lend confidence to asteroseismic masses in the metal poor regime. We
note that a larger sample will be needed to allow more stringent tests
to be made of systematic uncertainties in all the observables (both
seismic and non-seismic), {and to explore the properties of RHB stars, and of different populations in the
cluster}.

\end{abstract}


\section{Introduction}

Asteroseismology has revolutionised our view of evolved stars. The
NASA \textit{Kepler} \citep{Koch2010} and CNES-led CoRoT
\citep{Baglin2006} missions have delivered exquisite asteroseismic
data that have allowed radii and masses to be estimated for more than ten thousand
individual field red-giant stars in the Milky Way. These new results have direct implications for our ability
to determine distances and, crucially, to estimate ages of such stars,
which are key ingredients for in-depth studies of how the Galaxy
formed and evolved.


\begin{figure}
\centering
   \includegraphics[width=1\hsize]{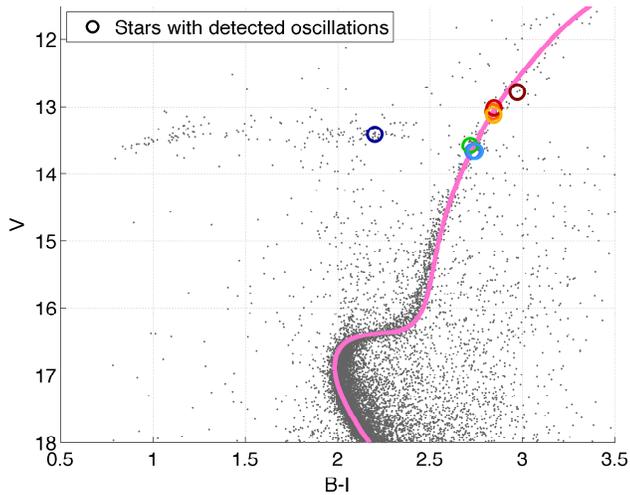}
      \caption{Colour magnitude diagram of M4 stars based on the dataset described in \citet{DAntona2009}. Magnitudes and colour were corrected for differential reddening following \citet{Milone2012}. The solid line represents a isochrone {(from BaSTI, see  Sec. \ref{sec:multiple} for details)} fit to the CMD. The
        large coloured open circles mark the stars with detected
        solar-like oscillations (the corresponding oscillation spectra
        are plotted in Fig. \ref{fig:PS}).}
         \label{fig:cmd}
\end{figure}


\begin{figure}
\centering
   \includegraphics[width=.9\hsize]{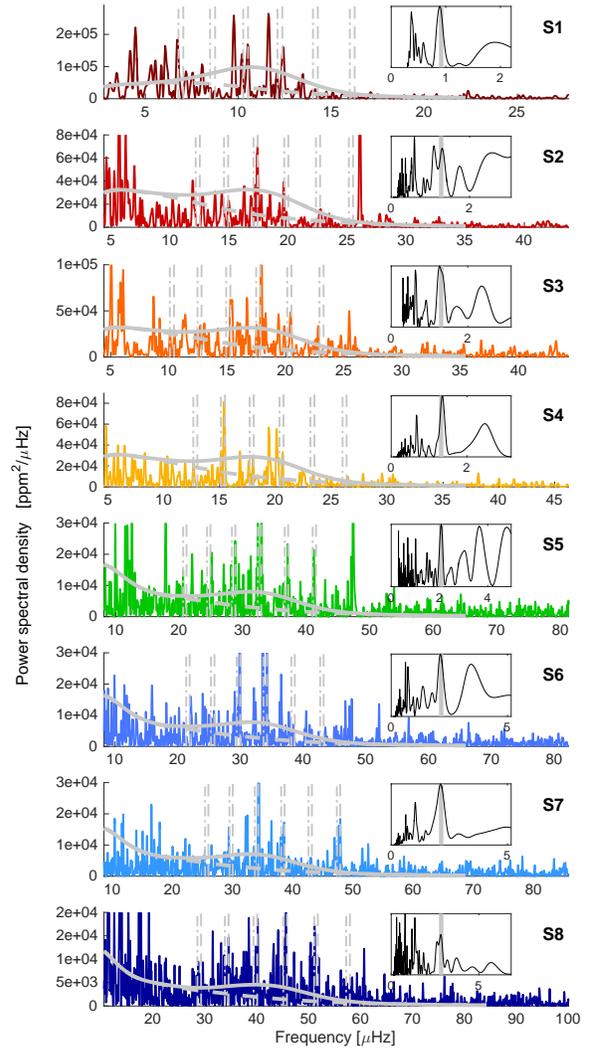}

      \caption{Solar-like oscillation spectra of eight K giants
        observed by K2. The bottom star is in the RHB. Stars (from top to bottom) are ordered by increasing $\nu_{\rm max}$.  Vertical lines show the position of radial (dashed lines) and quadrupole (dash-dotted lines) oscillation frequencies as expected by the pattern of oscillation modes in red-giant stars described in \citet{Mosser2011a}. {The expected granulation power and the combined granulation and oscillation power is represented by dashed and solid thick lines. The insets show the power spectrum of the power spectrum (PSPS) of each star, computed from the region around $\nu_{\rm max}$. In each PSPS the prominent peak at $\Delta\nu/2$ (vertical gray line) is the detected signature of the near-regular spacing of oscillation peaks in the frequency spectrum.}}
                 \label{fig:PS}
   \end{figure}
   

The strong correlation between mass and age in low-mass red-giant
stars means that the required goal of determining stellar ages to 30\% or
better implies that masses must be estimated to an accuracy better
than 10\%.  Comparisons against accurate and independent mass
determinations are however limited to stars in binary systems and,
most notably, stars in clusters \citep[see e.g.][for a
  review]{Brogaard2015}. Unfortunately, the open clusters observed by the 
\textit{Kepler} space telescope (during its nominal
mission) and by CoRoT explored the metal-rich regime only, and did not
provide a test of the metal-poor population.

Globular clusters are the oldest stellar systems for which it is
possible to make reliable age estimates, and are hence benchmarks to
test other age determinations. Previous efforts to detect solar-like oscillations
in globular clusters have been made, but either no oscillations were
detected \citep{Frandsen2007}, or only marginal detections were made
\citep{Stello2009}.  However, K2---the re-purposed \textit{Kepler}
mission \citep{Howell2014}---has now begun a survey of the ecliptic
plane, which contains bright clusters including the globular cluster
M4. In this Letter we report the detection of solar-like
oscillations in K2 data of K giants belonging to M4, and compare the measured
global oscillation properties against those expected from well-constrained, independent distance and mass estimates.

\section{M4 data reduction and analysis}

M4 was observed in K2 Campaign\,2 for a total of 78.8\,days. A
fraction of the cluster's total angular area on the celestial sphere
was covered by sixteen 50-by-50 pixel superstamps. Masks for
individual targets within the superstamps were defined using the $\rm
K2P^2$ \citep[K2-Pixel-Photometry;][]{Lund2015} pipeline. Each mask
was constructed from a summed image (over time) that allowed for the
apparent motion of the stars on the CCD due to the drift of the spacecraft  \citep[][]{Howell2014}.  Time-dependent positions
of stars on the CCD were estimated from the 2D cross-correlation of a
given superstamp, instead of estimated centroids for individual
targets. A set of unsupervised machine learning techniques was then
applied to define the final masks, from which lightcurves were then
produced.

Changes in measured flux due to spacecraft roll were corrected by
utilising the strong correlation of those changes with the stellar
position on the CCD, using procedures similar to those described in
\citet{Vanderburg2014}. The resulting, corrected lightcurves were then
cleaned for artifacts  
using the KASOC filter \citep{Handberg2014}; time scales of $\tau_{\rm long}=1$ days and $\tau_{\rm short}=0.25$ days were adopted for the median filters. We refer to \citet{Handberg2014} for additional details on the KASOC filter.

Among the sources identified by the K2P$^2$ pipeline we selected those
that could be unambiguously identified as K giants from the
\citet{DAntona2009} catalogue and the \citet{Marino2008,Marino2011}
membership studies. Moreover, we retained only stars with $V < 14$ and $
B-I > 1.7$, i.e. we avoided RR Lyrae pulsators, blue-horizontal-branch
stars, and stars that would be too faint to have detectable
oscillations (see Fig. \ref{fig:cmd}).

\begin{table*}
 \begin{tabular}{lcccccccccccc}
ID  & RA [deg]& DEC [deg] &2MASS ID &  $V$ & $V_{\rm dr}$ & $T_{\rm eff}$ [K] & $\Delta\nu$ [$\mu$Hz] & $\nu_{\rm max}$ [$\mu$Hz]  & Noise [ppm$^2$/$\mu$Hz] \\
\hline
S1& 245.850089&  -26.500147  &16232402-2630005   &    12.777  &    12.786  &   4585  &  $  1.83  \pm 0.02 $  & $  11.1    \pm   0.4  $  &  84\\
S2& 245.884870&  -26.439039&  16233236-2626205&      13.062 &   13.021   &  4715    &$  2.55  \pm  0.04 $ & $  17.2  \pm     0.7 $ & 211\\
S3& 245.911908&  -26.428539&  16233885-2625427&  13.071 &   13.071    & 4710   & $   2.62  \pm 0.04 $ &$   17.7     \pm  0.7 $ & 535\\
S4& 245.820426&  -26.496641&  16231690-2629479	&	13.096  &  13.121   &   4715    & $ 2.64  \pm     0.02  $ & $    18.5  \pm     0.7$ & 188\\
S5& 245.929534&  -26.468725&  16234308-2628074	&   13.539  &   13.583   & 4847   &  $4.14    \pm   0.02  $ & $    32.5  \pm     1.3 $ & 387 \\
S6& 245.949526&  -26.496729&  16234788-2629482	&   13.577  &   13.665   &  4842     & $ 4.30   \pm    0.02  $ & $    32.9  \pm     1.3 $& 202\\
S7& 245.841473&  -26.508892&  16232195-2630320	&   13.645  &   13.668   & 4805   &   $ 4.30 \pm 0.02 $ &  $ 34.3  \pm     1.4  $  & 172\\
S8& 245.985479&  -26.424564&  16235651-2625284	&    13.226  &   13.411  &   5672    & $ 5.67  \pm     0.05  $ & $    42.1  \pm     1.7$ & 192\\
\hline

 \end{tabular}
 \caption{Properties of the stars with detected solar-like oscillations. $V_{\rm dr}$ is the $V$-band magnitude from the dataset described in \citealt{DAntona2009}, corrected for differential reddening using the method described in \citet{Milone2012}. $T_{\rm eff}$ is calculated from corrected $B-V$ colour  and  \citet{Casagrande2014b}, the assumed uncertainty on $T_{\rm eff}$ is 100 K (see main text for details). } 
\label{tab:fullinfo}
 \end{table*}

{ 
We then searched the frequency-power spectra of the chosen set of {28}
lightcurves for evidence of solar-like oscillations using two independent detection pipelines. The first one is based on an updated
version (Elsworth et al., in preparation) of the automated detection
pipeline described in \citet{Hekker2010b} (see also
\citealt{Chaplin2015}). 
The asteroseismic analysis code was then used to extract from the
detected oscillation spectra estimates of two commonly used global or
average asteroseismic parameters: $\Delta\nu$, the average frequency
separation between consecutive overtones of modes having the same
angular degree; and $\nu_{\rm max}$, the frequency at which the
oscillations present their strongest observed amplitudes (see Elsworth et al., in preparation for details).
 {To compare the detected power with expectations we used the relations in \citet{Mosser2012} and \citet{Kallinger2014} to describe the power envelope due to the oscillations and the power spectrum of the granulation, to which we then added the contribution due to shot noise estimated from the mean power close to the Nyquist frequency (see Fig. \ref{fig:PS}). The observed power excess is compatible with expectations, in some cases weaker than expected, but this is in line with the fact that light curves of cluster stars suffer from a higher level of contamination from nearby sources}.

We have then performed a second analysis using  an independent method that effectively utilizes the expected frequency pattern of red giant oscillation modes \citep{Mosser2011a}. 
{Estimates of the large spacing were first provided by the autocorrelation function \citep{Mosser2009}, with the requirement that the null hypothesis be rejected at the  95\,\% confidence level.  These values were then refined with the method of \citet{Mosser2011a}. 
This uses a priori knowledge of the radial and quadrupole frequency patterns and provides reasonable constraints on the spacings even if the spectrum is of moderate quality only (see Fig. \ref{fig:PS} and \citealt{Hekker2012}).} Dipole modes, on the other hand,  are not used since their frequencies are expected to show a complex pattern originating from  the interaction between acoustic and gravity modes. 
{In all cases we also tested whether the excess power associated with a possible detection of the oscillations was consistent with expectations based on results from archival \emph{Kepler} data.}
For noisy spectra, the frequency of maximum oscillation was determined as for semi-regular variables showing only a limited number of modes \citep{Mosser2013}.

Having compared results, we retained only stars where both pipelines reported a detection of solar-like oscillations (8 stars, labelled S1 to S8; see Table \ref{tab:fullinfo} and Fig. \ref{fig:cmd}). Their power spectra
are shown in Fig. \ref{fig:PS}. Seven of the stars (S1 to S7) are on
the red-giant branch (RGB); the eighth (S8, spectrum shown in bottom
panel) is on the red horizontal branch (RHB).
We adopted  values and uncertainties for $\Delta\nu$ and $\nu_{\rm max}$ from the pipeline by \citet{Mosser2011a}, which are compatible within 1-$\sigma$ with those obtained with the first pipeline. {For two stars (S2 and S6) the first pipeline returned two possible solutions for $\Delta\nu$, while results from the  \citet{Mosser2011a} method returned only a single value (which was compatible with one of the two solutions of the first pipeline).}

Given the low fraction of stars in which we were able to unambiguously detect solar-like oscillations, we assessed the noise properties of the light curves analysed, and compared them with those of field stars.  The stars analysed in this work have a noise level (calculated as in \citealt{Stello2015}  as the median power between 260 and 280 $\mu$Hz) of the order of few hundreds ppm$^2/\mu$Hz {(see Table \ref{tab:fullinfo})}, which is {a factor $\sim$5-7} higher than in field stars of similar magnitude as presented in \citet{Stello2015}. A thorough assessment of whether the augmented noise is primarily due to the contamination from nearby sources in such a crowded field remains to be addressed. Moreover, tests  need to be carried out on how oscillation detection pipelines perform with K2 datasets, which are shorter, and have higher noise {(e.g. the instrumental noise peak at $\nu\simeq47.23$ $\mu$Hz, see \citealt{Lund2015})},  compared to those provided by the nominal \emph{Kepler} mission.}

\section{Results and comparison with independent constraints}
\label{sec:expect}

\subsection{Masses}
\label{sec:mass}

We proceeded as in \citet{Miglio2012} and estimated stellar masses by
using several combinations of the available seismic and non-seismic
constraints. 

{The average separation scales to very good approximation as the square root of the
mean density of the star, i.e., $\Delta\nu \propto \rho^{1/2}$; whilst
$\nu_{\rm max}$ has been found to scale with a combination of surface
gravity and effective temperature that also describes the dependence
of the cut-off frequency for acoustic waves in an isothermal
atmosphere, i.e., $\nu_{\rm max} \propto gT_{\rm eff}^{-1/2}$ (see
\citealt{Chaplin2013} for further details and references). }

Four sets of masses were computed, using:
 \begin{eqnarray}
 \frac{M}{{\rm M}_\odot} &\simeq& \left(\frac{\nu_{\rm max}}{\nu_{\rm
     max,
     \odot}}\right)^{3}\left(\frac{\Delta\nu}{\Delta\nu_{\odot}}\right)^{-4}\left(\frac{T_{\rm
     eff}}{{\rm T}_{\rm eff, \odot}}\right)^{3/2} \label{eq:scalM}, \\
 \frac{M}{{\rm M}_\odot} &\simeq&
 \left(\frac{\Delta\nu}{\Delta\nu_{\odot}}\right)^{2}\left(\frac{L}{{\rm
     L}_\odot}\right)^{3/2} \left(\frac{T_{\rm eff}}{{\rm T}_{\rm eff,
     \odot}}\right)^{-6} \label{eq:scalM2}, \\
 \frac{M}{{\rm M}_\odot} &\simeq& \left(\frac{\nu_{\rm max}}{\nu_{\rm
     max, \odot}}\right)\left(\frac{L}{{\rm
     L}_{\odot}}\right)\left(\frac{T_{\rm eff}}{{\rm T}_{\rm eff,
     \odot}}\right)^{-7/2} \label{eq:scalM3}, \\
 \frac{M}{{\rm M}_\odot} &\simeq& \left(\frac{\nu_{\rm max}}{\nu_{\rm
     max,
     \odot}}\right)^{12/5}\left(\frac{\Delta\nu}{\Delta\nu_{\odot}}\right)^{-14/5}\left(\frac{L}{{\rm
     L}_\odot}\right)^{3/10} {\rm .}\label{eq:scalM4}
 \end{eqnarray}
The solar reference values were taken as $\Delta\nu_\odot=135.1$
$\mu$Hz, $\nu_{\rm max,\odot}=3090$ $\mu$Hz \citep{Huber2013} and
${\rm T}_{\rm eff,\odot}=5777$ K. 
{The solar reference values for both pipelines used in this work differ from the values quoted by less than 0.5\%, hence the size of systematic shifts in mass when using Eq \ref{eq:scalM}-\ref{eq:scalM4} are expected to be lower than the uncertainties on the average RGB mass}.

The above equations assume strict
adherence to the classic asteroseismic scaling relations for
$\Delta\nu$ and $\nu_{\rm max}$.

Photometric $T_{\rm eff}$ were calculated using $(B-V)_{\rm 0}$ and compared with the value obtained using $(V-I)_{\rm 0}$ to check for consistency. We used $E(B-V)$ and $E(V-I)$ values from Table 3 in \citet{Hendricks2012}. Colour-$T_{\rm eff}$ calibrations, as well as bolometric correction (BC) at the stellar temperatures, and the solar BC were taken from \citet{Casagrande2014b}.
We iterated between the asteroseismic surface gravity, obtained from $\nu_{\rm max}$ and $T_{\rm eff}$, and the colour-$T_{\rm eff}$ relation, which requires the surface gravity as input.  In the colour-$T_{\rm eff}$ relation we  assumed the spectroscopically
determined $\rm [Fe/H]=-1.1$, and $\rm [\alpha/Fe]=0.4$, see \citealt{Marino2008}.  
We assumed an uncertainty on each $T_{\rm
  eff}$ of 100\,K.  We note that, for the 7 stars in common with the
analysis by \citet{Marino2008}, the spectroscopic and photometric
$T_{\rm eff}$ agree well within the uncertainties.

For internal consistency, the distance modulus was derived by
combining the {radii} of eclipsing binaries presented in
\citet{Kaluzny2013} with the temperatures from \citet{Casagrande2014b}, 
giving $(m-M)_{\rm 0}=11.20 \pm 0.10$. We then estimated stellar
luminosities using this distance together with the apparent
magnitudes, and bolometric corrections.

{For each set of masses from Eq. \ref{eq:scalM}--\ref{eq:scalM4}, formal uncertainties on the individual masses
were used to compute a weighted average mass of RGB stars ($\overline{M_{\rm RGB}}$). The
uncertainties in these averages were estimated from the weighted
scatter in the masses ($\sigma_{\overline{M}}$). To assess
how well the formal fitting uncertainties reflected the scatter in the
data we also report in Table {\ref{tab:meanmass}} the weighted mean uncertainty estimated
from the formal uncertainties on the masses
($\overline{\sigma_{\rm M}}$, see \citealt{Miglio2012} for details). 
In some cases
(Eq. \ref{eq:scalM2} and \ref{eq:scalM3}) the observed scatter is
significantly lower than expected from the formal uncertainties, which may indicate an overestimation of the observational uncertainties (e.g. on $T_{\rm eff}$ which have a significant systematic component and a high-power dependence in Eq. \ref{eq:scalM2} and \ref{eq:scalM3}).}

A source of possible systematic bias for masses determined using the
average or global asteroseismic parameters are known departures from
the classic scaling $\Delta\nu \propto \rho^{1/2}$ (see
e.g. discussions in \citealt{White2011, Miglio2012, Miglio2013,
  Belkacem2013}). Suggested corrections to the $\Delta\nu$ scaling are
likely to depend (at a level of few percent) on the stellar structure
itself.  To estimate a set of corrections we computed stellar models
using the code MESA \citep{Paxton2011}, taking an initial mass
$M=0.85\,\msun$ and heavy element abundance $Z=0.003$ (obtained using
the expression in \citealt{Salaris1993}, and the spectroscopically
determined metallicity
and alpha-enhancement {from \citealt{Marino2008}}). A Reimers'
mass-loss efficiency parameter of $\eta=0.2$ was also assumed.  For
any given model we defined $\Delta\nu$ to be a Gaussian-weighted
average {($\rm FWHM=0.66\,\nu_{\rm max}^{0.88}$, see \citealt{Mosser2012})}, 
centred in $\nu_{\rm max}$,  of the large frequency separations of adiabatic radial modes
 (for details see \citealt{Miglio2013} and
Rodrigues et al., in preparation). The $\Delta\nu$ values were normalised so
that a solar-calibrated model reproduced the average $\Delta\nu$
observed in the Sun.

Our results suggest that the seven RGB stars with detected
oscillations are in a $\nu_{\rm max}$ range where the mean density
will be underestimated by $\simeq 8\% $ when strict adherence of the
classic $\Delta\nu$ scaling is assumed.  For the RHB star the
comparison suggests an overestimation of the mean density by $\sim
4\%$.  If we apply these corrections to the mass determinations (see
last four rows of Table \ref{tab:meanmass}), we end up with a {significantly
lower scatter in the results (see also Fig.~\ref{fig:mass_id}) for all RGB stars}.

Needless
to say there are other sources of systematic uncertainty that may
affect the mass determination (e.g., systematic uncertainties on
$T_{\rm eff}$). A thorough description of the $\Delta\nu$ corrections, their
limitations and their dependences on stellar properties, will be
presented in Rodrigues et al., in preparation. 

Extracting individual mode frequencies from these data is
likely to be very challenging. Having estimates of individual 
frequencies,  and not just the
average $\Delta\nu$, would allow us to determine the stellar mean
density with a much improved precision (see e.g. \citealt{Huber2013},
Handberg et al. in preparation), and to mitigate the impact of our
poor modelling of surface layers \citep[e.g. see][]{Chaplin2013} and
of ambiguities in the definition of the average
$\Delta\nu$.

\subsubsection{Comparison with independent estimates of mass and distance}
\label{sec:multiple} 

\begin{figure}
\centering
   \includegraphics[width=1\hsize, angle=0]{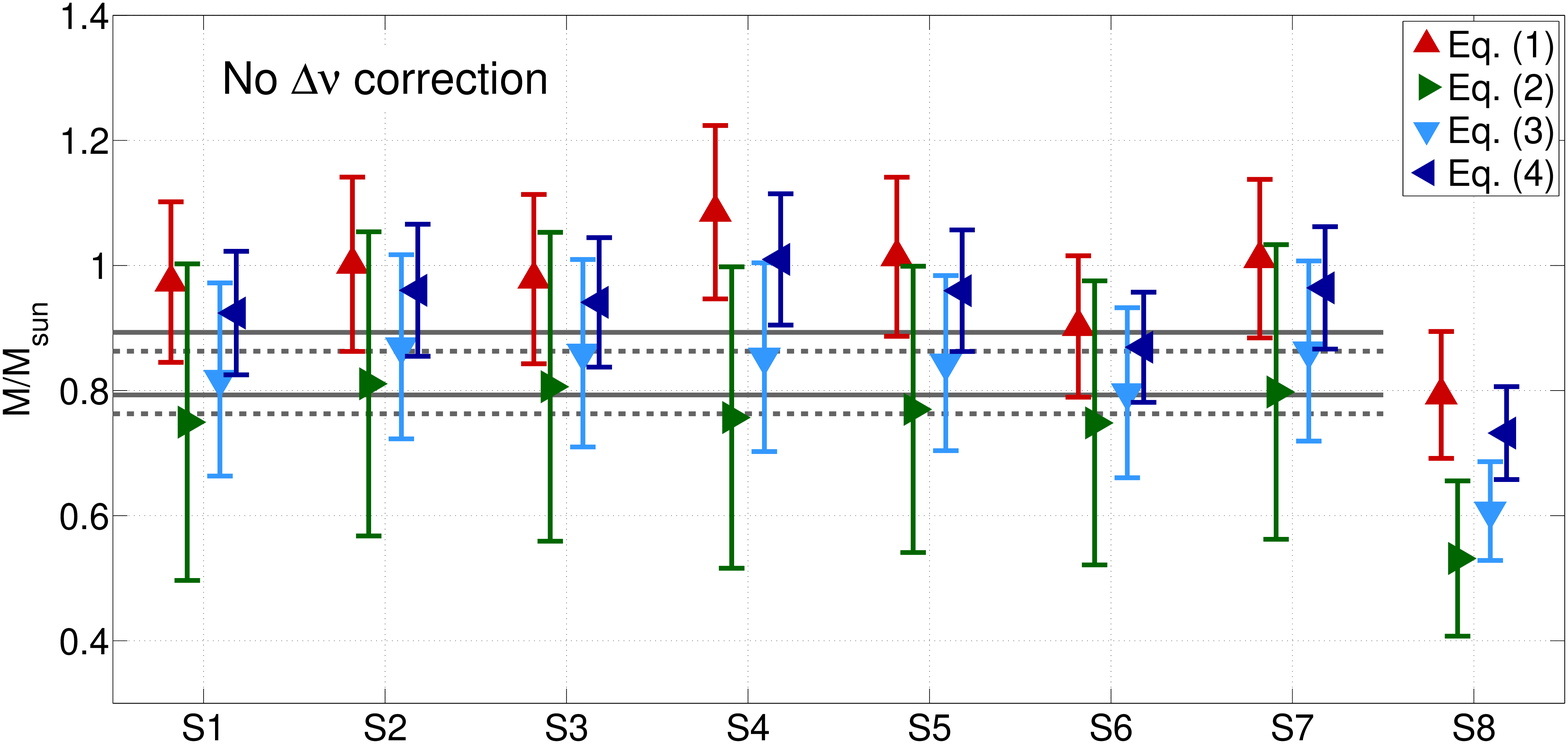}
      \includegraphics[width=1\hsize, angle=0]{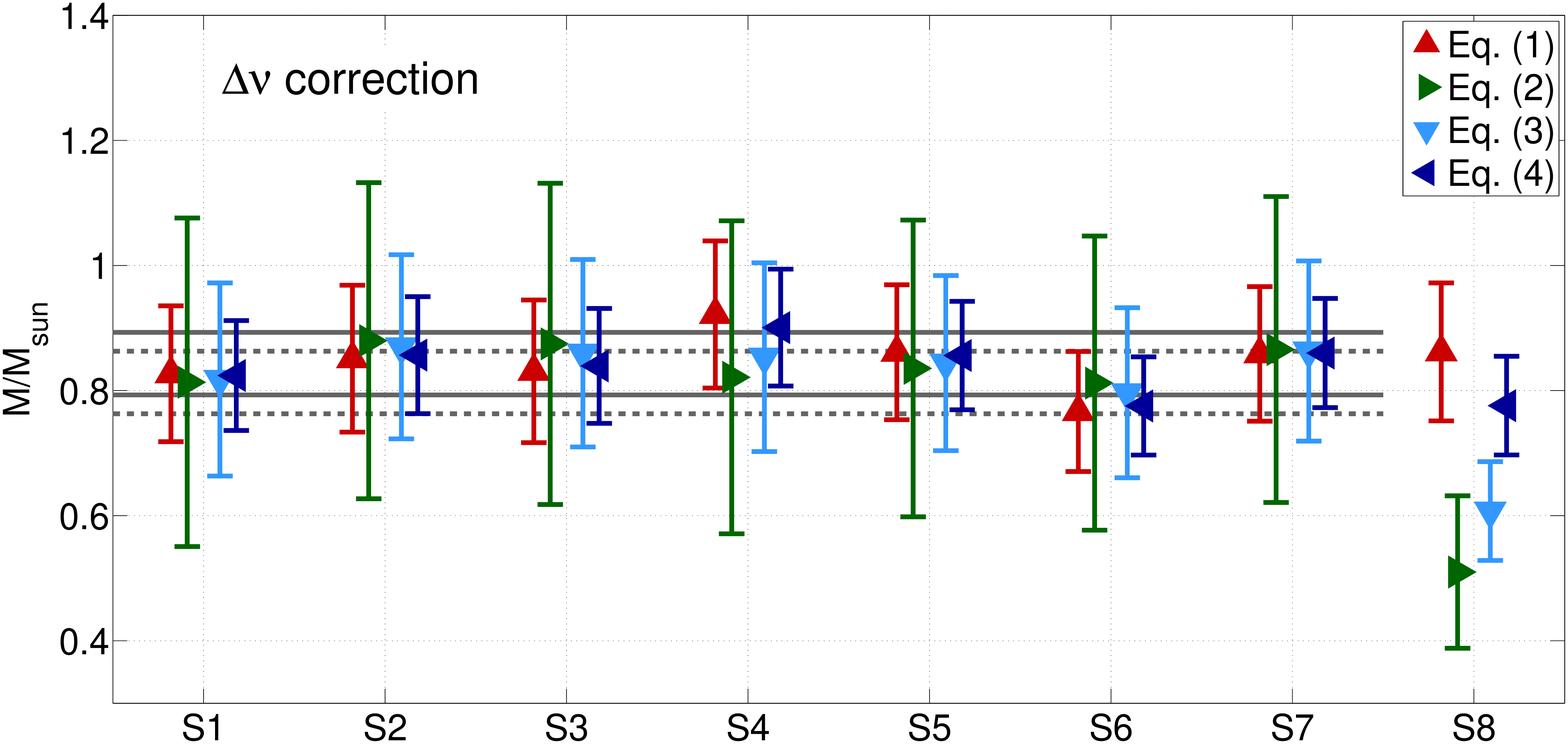}
      \caption{Mass of M4 giants as inferred from Eq. \ref{eq:scalM} to \ref{eq:scalM4} {with (lower panel) and without (upper panel) applying a model-predicted correction to the $\Delta\nu$ scaling relation}. The last star to the right (S8) is a RHB star. The solid and dashed lines denote the 1-$\sigma$ mass interval as determined from isochrone fitting and assuming two values for the initial He mass fraction (see Section \ref{sec:multiple} for details). }
         \label{fig:mass_id}
   \end{figure}

{ 
By fitting to the colour-magnitude diagram BaSTI
\citep{Pietrinferni2004} isochrones of the appropriate metallicity
and alpha-enhancement, and adopting an
initial He mass fraction $Y=0.25$, we find an age of 13 Gyr and  a $M_{\rm ISO, RGB}=0.84$ $\msun$.
We adopt a conservative uncertainty of $0.05$ $\msun${, which takes into account uncertainties on initial chemical composition, age, distance modulus and reddening}. This value
for $M_{\rm ISO, RGB}$ is also compatible with the value found by
extrapolating with isochrones the mass of the turnoff eclipsing
binaries \citep{Kaluzny2013} to the red giant branch phase{, which gives $M_{\rm EB, RGB}=0.85$ $\msun$.}

When the $\Delta\nu$ scaling is taken at face value,  $\overline{M_{\rm RGB}}$ determined from the four sets of masses, albeit
consistent ($\lesssim 10\%$) with the expected mass,  shows a significant scatter. When introducing a model-based correction to the $\Delta\nu$ scaling relation, the scatter between the various sets of masses is significantly reduced, and the discrepancy with independent mass estimates becomes smaller than the quoted uncertainties on the average mass ($\overline{\sigma_{\rm M}}$) and of the same order or smaller than the weighted scatter in the masses ($\sigma_{\overline{M}}$).  
A visual comparison between seismic masses and $M_{\rm ISO,RGB}$ is presented in
Fig.~\ref{fig:mass_id}.

The mass of the RHB star is marginally consistent with expectations ({$M_{\rm ISO, RHB} \simeq 0.74$ $\msun$}), and the model-suggested correction increases the scatter between the different mass estimates.  Given the uncertainty over mass-loss,  and hence on the expected correction to the $\Delta\nu$ scaling, increasing the number of RHB stars with detections will be crucial to quantify any significant bias in the seismic mass estimates.}

Several studies have revealed the existence of multiple stellar
populations, having different chemical compositions, in globular
clusters that have been subjected to a detailed abundance analysis
(e.g., see \citealt{Gratton2012, Piotto2015}).  M4 is no exception, and the
presence and properties of two main populations is well documented in the literature
\citep[see e.g.][]{Marino2008, Carretta2009, Milone2014, Malavolta2014}.  It is widely accepted
that the He-poor and He-rich populations in globular clusters are
coeval within a few hundreds Myr as predicted by the scenarios
proposed to explain the occurrence of these multiple populations
(e.g. see \citealt{Renzini2015} for a recent review on the proposed scenarios).

The present-day He-rich (Na-rich, O-poor) stars should therefore be
less massive than the He-poor (Na-poor, O-rich) stars because the former 
evolve more quickly.  The expected mass difference on the RGB
based on the different initial He mass fraction (0.25 versus 0.27; see
\citealt{Nardiello2015}) is inferred to be 0.03 $\rm M_{\odot}$ (using
BaSTI isochrones).  A higher He
enhancement, as suggested by, for example, \citet{Villanova2012},
would imply a higher mass difference \citep[see e.g.][for an
  exhaustive review on recent results]{Valcarce2014}.  The precision
in the average mass determined here is insufficient to detect this
difference.

 \begin{table}
 \centering

 \begin{tabular}{l | ccccc }
Eq. &  $\overline{M_{\rm RGB}}$ & $\overline{\sigma_{\rm M}}$ & $\sigma_{\overline{M}}$   & ${ N}$&  ${M_{\rm RHB}}$   \\
\hline
 (\ref{eq:scalM}) &      0.99  &  0.05     &0.02  &7 &  $ 0.79 \pm 0.10   $   \\
 (\ref{eq:scalM2}) &    0.78  &  0.09    & 0.01   &7 &   $ 0.53  \pm  0.12 $ \\
 (\ref{eq:scalM3}) &    0.84  &  0.06     &0.01  &7 &    $ 0.61 \pm   0.08  $ \\
 (\ref{eq:scalM4}) &    0.94  &  0.04     &0.02  &7&     $ 0.73  \pm  0.07 $\\
 \hline
 \multicolumn{6}{c}{$\Delta\nu_{\rm CORR}$}\\
\hline
(\ref{eq:scalM}) &    0.84   &0.04&     0.02 & 7 &  $  0.86 \pm   0.11   $   \\
 (\ref{eq:scalM2}) & 0.84    &0.09  &  0.01 & 7&   $  0.51 \pm   0.12  $ \\
 (\ref{eq:scalM3}) & 0.84    &0.06  &  0.01 & 7&    $  0.61  \pm  0.08   $ \\
 (\ref{eq:scalM4}) & 0.84    &0.03  &  0.02 & 7 &    $  0.78  \pm  0.08 $\\
\hline
 \end{tabular}

\caption{Average mass of stars on the RGB estimated using different
  observational constraints and scaling relations (Equations
  \ref{eq:scalM} to \ref{eq:scalM4}). $N$ is the number of stars
  included in the average. The masses reported in the last four
  rows were obtained introducing a correction to the $\Delta\nu$
  scaling as described in Sec. \ref{sec:mass}. The mass of the RHB star (S8) is reported in the last column.}

\label{tab:meanmass}

 \end{table}
 
\subsection{Radius / distance}

{Using a combination of seismic constraints and ${T_{\rm
     eff}}$,  we may also estimate}
stellar radii:
 \begin{equation}
 \frac{R}{{\rm R}_\odot} \simeq \left(\frac{\nu_{\rm max}}{\nu_{\rm
     max,
     \odot}}\right)\left(\frac{\Delta\nu}{\Delta\nu_{\odot}}\right)^{-2}\left(\frac{T_{\rm
     eff}}{{\rm T}_{\rm eff,
     \odot}}\right)^{1/2}{\rm.}\label{eq:scalR}
 \end{equation}
Radii determined from Eq. \ref{eq:scalR} agree at the 5\,\% level with
independent estimates determined from $L$ and \Teff. 

The above may also be formulated as a comparison of distance moduli.
{After applying the model-predicted correction to the $\Delta\nu$ scaling, }we find an average distance modulus (see Table \ref{tab:meandistance})
that is in excellent agreement with the independent determination obtained from constraints on eclipsing binaries  
 (see Sec. \ref{sec:mass}).

\begin{table}
\centering
 \begin{tabular}{ccccccc}
$\overline{DM}$ & $\sigma_{\rm \overline{DM}}$ & $\overline{\sigma_{\rm DM}} $ & $N$ & $\Delta\nu_{\rm CORR}$\\
\hline
 11.40 &     0.05  &   0.02   &8 & n\\
 11.26 &     0.05  &   0.06   &8& y\\

\hline
\end{tabular}
\caption{Mean true distance modulus ($DM=(m-M)_{\rm 0}$) and associated uncertainties, with and without introducing a correction to the $\Delta\nu$ scaling.}
\label{tab:meandistance}
\end{table}

\section{Summary and future prospects}

We have reported the first detections of solar-like oscillations in
giants belonging to a globular cluster. M4 provides what is at present
a unique set of targets for testing asteroseismic mass and radius
determination in low-metallicity environments. These tests are crucial
for the robustness of Galactic archeology studies, which are now
making use of solar-like oscillators \citep[see e.g.][]{Miglio2013a}.  In the sample of {RGB} stars analysed in our
study, we find no evidence for a significant systematic offset between
the seismic mass and radius/distance estimates and independent
determinations, {provided that a correction to the $\Delta\nu$ scaling relation as suggested by stellar models is introduced.  
In that case, for RGB stars we find   excellent agreement with the expected masses from isochrone fitting, and
using a distance modulus derived with independent methods. The offset with respect to independent masses is lower, or comparable with, the uncertainties on the average RGB mass ($4-10\%$, depending on the combination of constraints used).}

Extracting clean light curves from these crowded images is challenging, and further 
complicated by the instrumental drifts of K2. Having demonstrated that it is possible to 
detect solar-like oscillations in M4, we are now working on producing cleaner light curves for a larger
 sample of stars.
A systematic analysis
of asteroseismic detections in a larger sample of  M4 giants will
allow more stringent tests of the mass determination and, by
implication, systematic corrections to the asteroseismic $\Delta\nu$
scaling relation.

The detection of solar-like oscillations potentially opens the door to
the more ambitious goal of using seismology to probe multiple
populations in old globular clusters. Based on results in the literature
(\citealt{Marino2008} and \citealt{Carretta2009}), six of the M4 stars
with detected oscillations belong to the second (Na-rich, O-poor)
population, while the RHB star and S2 are likely to be first
generation (Na-poor, O-rich) stars. Again, an increase in the number
of stars with detections of solar-like oscillations may allow us to detect mass differences
between multiple populations, although the systematic uncertainties
described in Section \ref{sec:expect} will need to be borne in mind.

Looking to the future, neither the upcoming NASA TESS Mission
\citep{Ricker2014} nor the ESA PLATO Mission \citep{Rauer2014} are
optimized for the study of densely populated stellar clusters.  A
space mission dedicated to the detection and study of oscillations in
globular clusters should be considered. Long-duration observations,
like the multi-year observations provided by the nominal \emph{Kepler}
mission, would give the frequency resolution needed to extract
individual frequencies of many modes. This would not only improve the
determination of global properties (see Section \ref{sec:mass}) but
also give us access to seismic proxies of the internal structures of
the stars (i.e., the near-core structure, internal rotation, and
information on the envelope He abundance).  The limitations imposed by
the shorter-duration campaigns of K2 mean that extracting individual
frequencies of red giants from the existing M4 data will be much more
challenging.

\section*{Acknowledgments}

Funding for this Discovery mission is provided by NASA's Science
Mission Directorate. The authors wish to thank the entire {\it Kepler} and K2 
team, without whom these results would not be possible. A.M.,
W.J.C., G.R.D., Y.P.E., T.C., C.J., and J.S.K. acknowledge the support of the UK Science and Technology
Facilities Council (STFC).  Funding for the Stellar Astrophysics Centre
is provided by The Danish National Research Foundation (Grant
agreement no.: DNRF106). The research is supported by the ASTERISK
project (ASTERoseismic Investigations with SONG and Kepler) funded by
the European Research Council (Grant agreement no.: 267864). M.N.L. acknowledges the support of The Danish Council for Independent Research | Natural Science (Grant DFF-4181-00415), and the European Community's Seventh Framework Programme (FP7/2007-2013) under grant agreement no. 312844 (SPACEINN).  APM and AFM acknowledge support by the Australian Research Council through Discovery Early Career Researcher Awards DE150101816 and DE160100851.
L.R.B.,  L.M. and V.N. acknowledge PRIN-INAF 2012 funding under the project
entitled: 'The M4 Core Project with Hubble Space Telescope'. L.M. acknowledges the financial support from the European Union Seventh Framework Programme (FP7/2007-2013) under Grant agreement number 313014 (ETAEARTH).

\bibliographystyle{mn2e_new}
\small
\bibliography{andrea_m}
\label{lastpage}
\end{document}